\documentclass[12pt,preprint]{emulateapj}
\usepackage{graphicx}

\shorttitle{The HD 99492 System}
\shortauthors{Stephen R. Kane et al.}
\slugcomment{Submitted for publication in the Astrophysical Journal
  Letters}

\begin{document}

\title{Stellar Activity and Exclusion of the Outer Planet in the HD
  99492 System}

\author{
  Stephen R. Kane\altaffilmark{1},
  Badrinath Thirumalachari\altaffilmark{1},
  Gregory W. Henry\altaffilmark{2},
  Natalie R. Hinkel\altaffilmark{1,3},
  Eric L.N. Jensen\altaffilmark{4},
  Tabetha S. Boyajian\altaffilmark{5},
  Debra A. Fischer\altaffilmark{5},
  Andrew W. Howard\altaffilmark{6},
  Howard T. Isaacson\altaffilmark{7},
  Jason T. Wright\altaffilmark{8,9}
}
\email{skane@sfsu.edu}
\altaffiltext{1}{Department of Physics \& Astronomy, San Francisco
  State University, 1600 Holloway Avenue, San Francisco, CA 94132,
  USA}
\altaffiltext{2}{Center of Excellence in Information Systems, Tennessee
  State University, 3500 John A. Merritt Blvd., Box 9501, Nashville,
  TN 37209, USA}
\altaffiltext{3}{School of Earth \& Space Exploration, Arizona State
  University, Tempe, AZ 85287, USA}
\altaffiltext{4}{Dept of Physics \& Astronomy, Swarthmore College,
  Swarthmore, PA 19081, USA}
\altaffiltext{5}{Department of Astronomy, Yale University, New Haven,
  CT 06511, USA}
\altaffiltext{6}{Institute for Astronomy, University of Hawaii,
  Honolulu, HI 96822, USA}
\altaffiltext{7}{Astronomy Department, University of California,
  Berkeley, CA 94720, USA}
\altaffiltext{8}{Department of Astronomy and Astrophysics,
  Pennsylvania State University, 525 Davey Laboratory, University
  Park, PA 16802, USA}
\altaffiltext{9}{Center for Exoplanets \& Habitable Worlds,
  Pennsylvania State University, 525 Davey Laboratory, University
  Park, PA 16802, USA}


\begin{abstract}

A historical problem for indirect exoplanet detection has been
contending with the intrinsic variability of the host star. If the
variability is periodic, it can easily mimic various exoplanet
signatures, such as radial velocity variations that originate with the
stellar surface rather than the presence of a planet. Here we present
an update for the HD~99492 planetary system, using new radial velocity
and photometric measurements from the Transit Ephemeris Refinement and
Monitoring Survey (TERMS). Our extended time series and subsequent
analyses of the Ca II H\&K emission lines show that the host star has
an activity cycle of $\sim$13 years. The activity cycle correlates
with the purported orbital period of the outer planet, the signature
of which is thus likely due to the host star activity. We further
include a revised Keplerian orbital solution for the remaining planet,
along with a new transit ephemeris. Our transit-search observations
were inconclusive.

\end{abstract}



\section{Introduction}
\label{introduction}

The radial velocity (RV) technique remains one of the most successful
methods for the discovery of exoplanetary systems. At the present
time, more than 500 exoplanets have been discovered using the RV
technique, including a vast range of multi-planet systems and orbital
configurations. The success of this method is greatly dependent upon
the ability to accurately characterize the properties of the host
star. In particular, the evolution of star spots, magnetic fields, and
pulsations have well-studied effects on stellar radial velocity
variations \citep{saa97,que01,des07,heb14}. There have been numerous
recent cases where stellar activity has posed a significant problem in
the correct interpretation of RV data \citep{hat13,hat15,rob14,rob15}.

One source of activity-induced RV variations is that due to stellar
activity cycles, analogous to the 11-year Solar cycle. \citet{dra85}
predicted such a correlation, and \citet{dem87} reported the detection
of such a correlation in the solar CO lines at 2.3$\mu$m, and inferred
an amplitude of 30~m\,s$^{-1}$ from the effect. \citet{wri08} argued
that experience with the hundreds of sun-like stars from the
California Planet Survey (CPS) showed that such effects are not so
strong, and that activity cycles were probably not to blame for a
$\sim$15~m\,s$^{-1}$ RV variation in phase with an activity cycle in
HD~154345. Similar high-amplitude RV-activity correlations in
individual targets have been reported by \citet{mou11}, \citet{car14},
and \citet{rob13}. Nonetheless, for most stars such correlations are
small or absent, as argued by \citet{wri08} and \citet{san10}.
 
The star HD~99492 is an early-K dwarf in a binary orbit with HD~99491
(also known as 83 Leonis B and A, respectively). HD~99492 has a
parallax of $55.7\pm1.46$~marcsec and a distance of $17.96\pm0.47$~pcs
\citep{van07a, van07b}. The mean angular separation of the stellar
components is 40.76\arcsec, leading to an average projected separation
of $\sim$730~AU. HD~99492 was found to harbor a 0.1~$M_J$ planet in a
17 day orbit by \citet{mar05}. The best-fit Keplerian orbital solution
at that time included a linear trend to account for a possible second
companion in the system. The orbital elements were updated by
\citet{mes11}, who claimed to have resolved the separate orbit of an
outer planet with a period of $\sim$5000 days.

\begin{figure*}
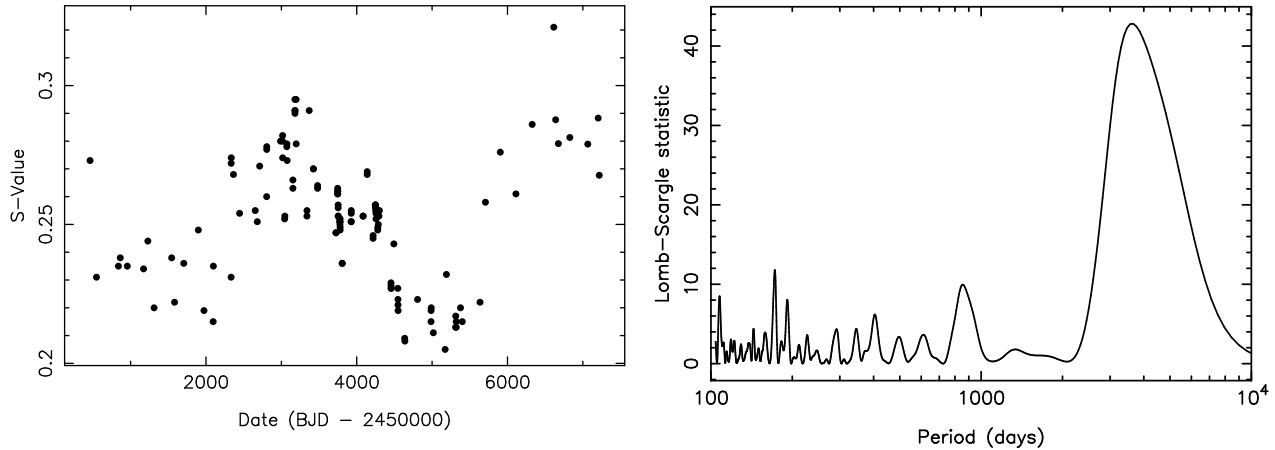

  \begin{center}
    \begin{tabular}{cc}
      \includegraphics[angle=270,width=8.2cm]{f01a.ps} &
      \includegraphics[angle=270,width=8.2cm]{f01b.ps}
    \end{tabular}
  \end{center}
  \caption{{\it Left}: HD~99492 S-values determined from the complete
    time series of Keck/HIRES spectra. {\it Right:} The periodogram
    resulting from a fourier analysis of the HD~99492 S-values,
    revealing a broad peak between 3000--5000~days.}
  \label{actfig}
\end{figure*}

Here we present new results for the system that reveal an activity
cycle in the star and further show that stellar activity amply
explains the signature of the outer planet (c). Section \ref{stellar}
provides new fundamental stellar parameters, including spectral
analysis, discussion of element abundances, and activity indices from
the complete dataset of 130 Keck/HIRES spectra. Section \ref{update}
presents our revised Keplerian orbital solution, including the
correlation of the outer planet signature with the activity
indices. Section \ref{photometry} includes photometry from 5 observing
seasons acquired over a span of 11 years. The photometric data
confirms the absence of brightness variations in phase with the
orbital period of planet b, thus confirming the radial velocity
variations in HD 99492 on a 17 day cycle are due to planetary reflex
motion. Our limited number of brightness measurements near the
predicted phase of planetary transit show no evidence for a transit
but fall short of ruling them out. We provide concluding remarks in
Section~\ref{conclusions}.


\section{Stellar Properties}
\label{stellar}


\subsection{Fundamental Parameters}
\label{stellar:sme}

The fundamental properties of HD~99492 have been previously
determined, for example by \citet{val05,tak07}. We used an upgraded
version of the Spectroscopy Made Easy (SME) package to model a
Keck/HIRES spectrum of HD~99492. Details of the SME package may be
found in \citet{val96,val05}. Briefly, SME uses an iterative technique
that combines model atmosphere analysis with Yonsei-Yale model
isochrones \citep{dem04} that utilize {\it Hipparcos} photometry and
distances \citep{van07a,van07b}. This approach produces a
self-consistent convergence with the measured surface gravity
\citep{val09}.

The results of our analysis are shown in Table~\ref{system}, including
values for the surface gravity $\log g$, rotational velocity $v \sin
i$, atmospheric abundance [Fe/H], effective temperature $T_{\rm eff}$
and stellar isochrone solution (mass $M_\star$, radius $R_\star$, and
age). These parameters are consistent with previous estimates of the
stellar properties and demonstrate that HD~99492 is a late-G/early-K
dwarf with an age similar to the Sun.


\subsection{Stellar Abundances}
\label{sec:abund}

The element abundances of HD~99492 have been measured only by two
groups to-date, namely \citet{val05} and \citet{pet11}. To correct for
varying solar abundance normalizations, per the analysis within the
Hypatia Catalog \citep{hin14}, each dataset was re-normalized to the
\citet{lod09} scale. The [Fe/H] measurement per both groups is 0.40
dex, since \citet{pet11} adopted the stellar parameters and iron
abundance from \citet{val05} in their analysis. From \citet{pet11},
[O/H] $=$ 0.25 dex while \citet{val05} determined [Na/H] $=$ 0.41,
[Si/H] $=$ 0.34 dex, [Ti/H] $=$ 0.28 dex, and [Ni/H] $=$ 0.38 dex.
These results reveal a star that is markedly super-solar in both the
volatile and refractory elements.
  

\subsection{Stellar Activity}
\label{activity}

HD~99492 has been spectroscopically monitored using the HIRES echelle
spectrometer \citep{vog94} on the 10.0m Keck I telescope as part of
the CPS. For Keck/HIRES instrument configuration details, see
\citet{wri04,how09}. Our complete HIRES dataset contains 130
measurements spanning over 18 years, extending the time baseline of
the data reported by \citet{mes11} by over 5 years. The pipeline that
extracts the RVs from the spectra (see Section~\ref{update}) also
extracts Ca II H\&K line-profile variations and provides an index of
stellar activity \citep{noy84}. These data are calibrated to the
Mt. Wilson S-values, defined as the ratio of the sum of the flux in
the H\&K line cores to the sum of the two continuum bands on either
side \citep{wil68}. We include data acquired both before and after the
upgrade of the HIRES CCD in 2004 August \citep{isa10}, taking into
account the offset between pre-2004 and post-2004 calibrated datasets.

The time series of S-values are shown in the left panel of
Figure~\ref{actfig}. The periodic variation in S indicates that we
have observed just over one complete cycle of stellar activity in the
host star. To quantify the variation, we performed a fourier analysis
of the time series, resulting in the periodogram shown in the right
panel of Figure~\ref{actfig}. This analysis reveals a broad peak in
the power spectrum that lies between 3000--5000 days, with maximum
power occurring at $\sim$3650~days. The S-value periodicity is thus
consistent with the HD~99492c orbital period of $4970\pm744$~days
determined by \citet{mes11}. We elaborate further on the correlation
between stellar activity and possible planetary signature in
Section~\ref{update}.


\section{An Update to the Planetary System}
\label{update}

\begin{figure}
  \includegraphics[width=8.2cm]{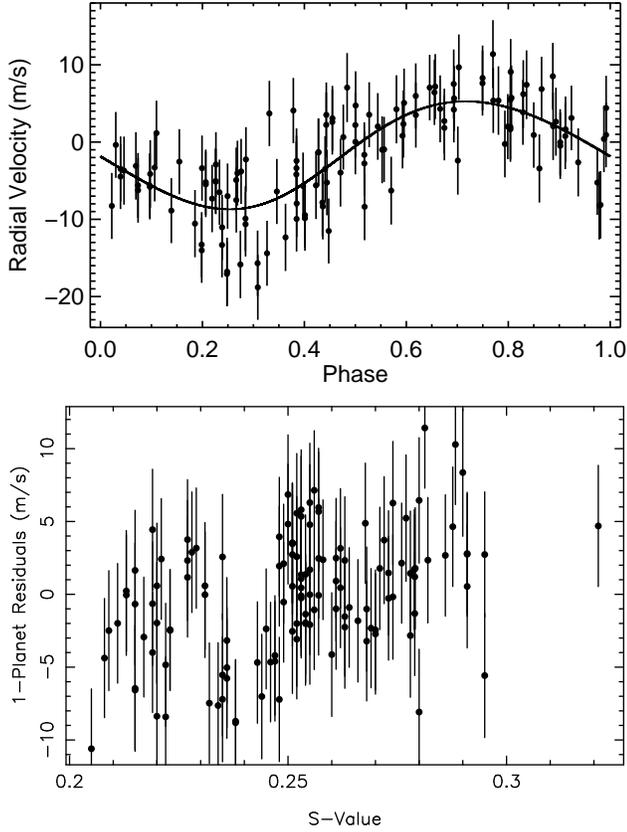} \\
  \includegraphics[angle=270,width=8.2cm]{f02b.ps}
  \caption{{\it Top}: The complete 130 RV measurement dataset phased
    on the best-fit Keplerian orbital solution for a single-planet
    system (see Table~\ref{system}). {\it Bottom:} The residuals from
    the best-fit solution plotted against the activity indices
    described in Section~\ref{activity}. Our analysis shows that the
    probability of no correlation between the one-planet RV residuals and
    the S-values is $1.2 \times 10^{-5}$.}
  \label{rvfig}
\end{figure}

The RV measurements were extracted from the Keck/HIRES data with the
use of an iodine cell mounted at the spectrometer entrance slit as a
robust source of wavelength calibration \citep{mar92,val95}. The
modeling procedure for the Doppler shift of each stellar spectrum with
respect to the iodine spectrum is described further in
\citet{how09}. The discovery orbital solution for the HD~99492 system
by \citet{mar05} included a linear trend component. The 93 RV
measurements utilized by \citet{mes11} used a two-planet orbital
solution to account for the previously-noted linear trend. A
two-planet fit to our expanded dataset is able to recover a similar
orbital solution to that previously found by
\citet{mes11}. Considering the periodic stellar activity described in
Section~\ref{activity} as a source for the previously observed linear
trend and purported second planet, we performed a single-planet fit to
our dataset of 130 RV measurements, both with and without a linear
trend included. These fits were carried out using RVLIN, a
partially-linearized, least-squares fitting procedure described in
\citet{wri09}. The uncertainties in the resulting orbital parameters
were estimated using the BOOTTRAN bootstrapping routines described in
\citet{wan12}. We included a stellar jitter noise component of
4~m\,s$^{-1}$ in quadrature with the measurement uncertainties
\citep{wri05,but06}. With our new dataset and its increased timespan,
we find no evidence of a significant difference between the orbital
fits that do and do not include a linear trend. We thus adopt the
solution without the linear trend for which the complete orbital
solution is shown in Table~\ref{system} and in the top panel of Figure
\ref{rvfig}. Note that the $\gamma$ parameter shown in
Table~\ref{system} is the systemic velocity of the system with respect
to the zero point of the extracted RVs and thus is the systemic
velocity relative to the template spectrum. The complete RV dataset of
130 measurements of HD~99492 are listed in Table~\ref{rvs}.
  
\begin{deluxetable}{lc}
  \tablecaption{\label{system} System Parameters}
  \tablewidth{0pt}
  \tablehead{
    \colhead{Parameter} &
    \colhead{Value}
  }
  \startdata
\noalign{\vskip -3pt}
\sidehead{HD99492}
~~~~$V$                           & 7.58 \\
~~~~$B-V$                         & 1.0 \\
~~~~Distance (pc)                 & $55.7 \pm 1.46$ \\
~~~~$T_\mathrm{eff}$ (K)          & $4929 \pm 44$ \\
~~~~$\log g$                      & $4.57 \pm 0.06$ \\
~~~~$v \sin i$ (km\,s$^{-1}$)     & $0.41 \pm 0.5$ \\
~~~~$[$Fe/H$]$ (dex)              & $0.3 \pm 0.03$ \\
~~~~$M_\star$ ($M_\odot$)         & $0.85 \pm 0.02$ \\
~~~~$R_\star$ ($R_\odot$)         & $0.78 \pm 0.02$ \\
~~~~Age (Gyrs)                    & $4.8 \pm 4.1$ \\
\sidehead{HD 99492 b}
~~~~$P$ (days)                    & $17.054 \pm 0.003$ \\
~~~~$T_c\,^{a}$ (JD -- 2,440,000) & $17367.776 \pm 0.855$ \\
~~~~$T_p\,^{b}$ (JD -- 2,440,000) & $13776.317 \pm 3.392$ \\
~~~~$e$                           & $0.07 \pm 0.06$ \\
~~~~$\omega$ (deg)                & $240.7 \pm 75.4$ \\
~~~~$K$ (m\,s$^{-1}$)             & $6.98 \pm 0.53$ \\
~~~~$M_p$\,sin\,$i$ ($M_J$)       & $0.079 \pm 0.006$ \\
~~~~$a$ (AU)                      & $0.123 \pm 0.001$ \\
\sidehead{System Properties}
~~~~$\gamma$ (m\,s$^{-1}$)           & $-1.49 \pm 0.37$ \\
\sidehead{Measurements and Model}
~~~~$N_{\mathrm{obs}}$            & 130 \\
~~~~rms (m\,s$^{-1}$)             & 4.33 \\
~~~~$\chi^2_{\mathrm{red}}$       & 1.03
  \enddata
  \tablenotetext{a}{Time of mid-transit.}
  \tablenotetext{b}{Time of periastron passage.}
\end{deluxetable}

To investigate further the impact of stellar activity on a two-planet
solution (see Section~\ref{activity}), we compared the S-values with
the RV residuals of the single-planet solution shown in
Table~\ref{system}. The resulting correlation diagram is shown in the
bottom panel of Figure~\ref{rvfig}. We quantified the significance of
the correlation using the Spearman rank correlation coefficient. The
Spearman coefficient lies in the range $-1 < r_s < 1$, and, in turn,
gives the probability that the two quantities being examined are not
correlated. The Spearman coefficient for the data shown in the bottom
panel of Figure~\ref{rvfig} is $r_s = 0.39$, indicative of a positive
correlation. The corresponding probability that the residuals of the
single-planet solution and the S-values would produce the observed
correlation if those quantities were in fact uncorrelated is $1.2
\times 10^{-5}$. We conducted a further test via an extensive
Monte-Carlo simulation that performs a Fisher-Yates shuffle,
randomizing the order of the residual data values. For each
realization, the Spearman's rank correlation coefficient and
probability were recalculated. This test resulted in a 0.5 probability
of null-correlation, indicating that the correlation found above is
robust. This implies, in turn, that the second planet claimed by
\citet{mes11} is instead the result of stellar activity.


\section{Photometric Observations}
\label{photometry}

We observed HD~99492 photometrically as part of the Transit Ephemeris
Refinement and Monitoring Survey (TERMS) \citep{kan09} with the T12
0.8m Automatic Photoelectric Telescope (APT), one of several automated
telescopes operated by Tennessee State University (TSU) at Fairborn
Observatory in southern Arizona. The T12 APT is equipped with a
precision, two-channel photometer that simultaneously measures the
Str\"omgren $b$ and $y$ passbands using two EMI 9924QB photomultiplier
tubes. This makes T12 ideal for achieving high photometric precision
on relatively bright stars. The TSU APTs and their precision
photometers, observing strategy, data reduction techniques, and
photometric precision are described in detail by \citet{hen99}.

\LongTables
\begin{deluxetable}{ccc}
  \tablewidth{0pc}
  \tablecaption{\label{rvs} HD~99492 Radial Velocities}
  \tablehead{
    \colhead{Date} &
    \colhead{RV} &
    \colhead{$\sigma$} \\
    \colhead{(BJD -- 2,440,000)} &
    \colhead{(m\,s$^{-1}$)} &
    \colhead{(m\,s$^{-1}$)}
  }
  \startdata
10462.113958  &  -3.09  & 1.49  \\
10546.987859  &  -3.70  & 1.62  \\
10837.932535  &  -3.30  & 1.56  \\
10862.898993  &  -6.29  & 1.57  \\
10955.876644  &  -8.27  & 1.21  \\
11172.101597  &  -2.40  & 1.62  \\
11228.035903  &  -8.13  & 1.51  \\
11311.816319  &  2.63  & 1.59  \\
11544.172650  &  -8.38  & 1.39  \\
11582.974942  &  -0.25  & 1.35  \\
11704.805914  &  -2.63  & 1.62  \\
11898.154005  &  -15.85  & 1.48  \\
11973.053090  &  4.30  & 1.39  \\
12095.752049  &  -3.42  & 1.63  \\
12097.753715  &  -8.23  & 1.56  \\
12333.139410  &  5.37  & 1.69  \\
12334.079884  &  7.42  & 1.73  \\
12334.968322  &  2.05  & 1.53  \\
12364.068125  &  2.33  & 1.46  \\
12445.768264  &  -7.95  & 1.40  \\
12654.009595  &  5.07  & 1.70  \\
12681.123484  &  -10.56  & 1.46  \\
12711.858843  &  0.40  & 1.28  \\
12804.765590  &  -7.83  & 1.43  \\
12805.876296  &  4.73  & 1.68  \\
12806.763634  &  -1.00  & 1.39  \\
12989.171424  &  -16.78  & 1.60  \\
13015.119444  &  11.38  & 1.59  \\
13016.134363  &  6.20  & 1.59  \\
13017.121921  &  8.51  & 1.38  \\
13044.127569  &  -3.96  & 1.58  \\
13045.999074  &  4.24  & 1.62  \\
13071.870764  &  -4.13  & 1.51  \\
13073.940752  &  -7.32  & 1.47  \\
13076.983611  &  -5.70  & 1.42  \\
13153.801470  &  -0.04  & 1.29  \\
13153.804144  &  -0.48  & 1.28  \\
13179.820787  &  -1.32  & 1.46  \\
13179.824352  &  -1.33  & 1.57  \\
13180.782037  &  7.05  & 1.78  \\
13181.808171  &  2.02  & 1.35  \\
13195.775914  &  -12.34  & 1.46  \\
13196.794780  &  -5.61  & 1.47  \\
13339.157731  &  5.37  & 0.89  \\
13340.150718  &  3.85  & 0.95  \\
13369.115093  &  3.54  & 0.92  \\
13425.000741  &  1.95  & 0.97  \\
13425.003310  &  1.67  & 1.02  \\
13480.759734  &  -5.61  & 0.92  \\
13480.761887  &  -6.24  & 0.90  \\
13725.101748  &  -9.87  & 0.93  \\
13725.104595  &  -9.46  & 0.91  \\
13747.133588  &  7.52  & 1.77  \\
13747.138218  &  4.20  & 1.10  \\
13747.145174  &  5.65  & 0.98  \\
13748.096169  &  7.62  & 0.91  \\
13748.098819  &  8.29  & 0.92  \\
13753.040035  &  -4.47  & 0.91  \\
13753.043380  &  -3.48  & 0.92  \\
13754.021562  &  -5.64  & 0.92  \\
13754.024097  &  -5.80  & 0.96  \\
13775.980868  &  -3.38  & 0.96  \\
13775.983125  &  -2.44  & 0.93  \\
13776.976910  &  2.23  & 0.91  \\
13776.979213  &  3.52  & 0.90  \\
13777.950347  &  2.20  & 0.95  \\
13777.952720  &  0.03  & 0.94  \\
13779.971238  &  5.97  & 0.99  \\
13779.974155  &  3.50  & 1.00  \\
13806.916794  &  -13.28  & 0.95  \\
13806.918981  &  -14.02  & 0.95  \\
13926.762188  &  -5.05  & 0.99  \\
13926.768762  &  -5.12  & 0.98  \\
13927.761840  &  -9.91  & 0.90  \\
13927.764213  &  -10.64  & 0.82  \\
14084.153623  &  2.60  & 1.02  \\
14084.157870  &  3.07  & 0.94  \\
14139.063102  &  2.72  & 0.86  \\
14139.064722  &  1.83  & 0.92  \\
14216.896134  &  -13.32  & 0.91  \\
14216.899722  &  -11.04  & 0.99  \\
14246.798900  &  0.92  & 0.85  \\
14246.800718  &  4.41  & 0.75  \\
14248.811678  &  1.16  & 0.87  \\
14250.800613  &  -2.91  & 0.84  \\
14251.804815  &  -2.25  & 0.82  \\
14255.765556  &  -1.65  & 0.89  \\
14255.766991  &  -2.77  & 0.90  \\
14277.743067  &  5.72  & 0.85  \\
14278.749942  &  6.86  & 0.83  \\
14279.748507  &  3.12  & 0.82  \\
14285.751910  &  -3.80  & 0.97  \\
14294.758669  &  5.47  & 0.97  \\
14300.738970  &  -2.54  & 0.96  \\
14455.109028  &  -5.20  & 1.04  \\
14455.110868  &  -5.48  & 1.06  \\
14456.129444  &  -4.92  & 0.96  \\
14456.131400  &  -7.51  & 0.92  \\
14493.134583  &  -8.34  & 1.17  \\
14544.982280  &  0.64  & 0.95  \\
14546.963137  &  0.81  & 1.03  \\
14547.871944  &  7.05  & 1.10  \\
14548.847407  &  9.68  & 1.10  \\
14635.754444  &  2.00  & 0.98  \\
14638.750949  &  -5.25  & 0.86  \\
14807.164861  &  0.91  & 1.08  \\
14985.837363  &  -14.40  & 0.93  \\
14986.825959  &  -9.95  & 1.00  \\
14987.839171  &  -5.24  & 1.08  \\
15016.744103  &  -8.88  & 0.93  \\
15173.123927  &  -18.81  & 0.94  \\
15190.171113  &  -15.69  & 1.05  \\
15311.807895  &  -3.49  & 1.09  \\
15313.781452  &  -0.95  & 0.97  \\
15319.842967  &  0.77  & 1.23  \\
15319.850617  &  1.65  & 1.11  \\
15376.739902  &  -17.07  & 0.89  \\
15400.735697  &  6.43  & 1.01  \\
15635.955861  &  -11.51  & 0.94  \\
15707.736240  &  7.18  & 0.95  \\
15905.166211  &  -6.50  & 1.01  \\
16111.736847  &  -6.40  & 0.95  \\
16328.051766  &  -0.37  & 1.17  \\
16614.127514  &  9.08  & 1.15  \\
16639.094368  &  -4.03  & 0.91  \\
16675.173277  &  -4.19  & 1.18  \\
16827.757817  &  3.70  & 1.03  \\
17065.116057  &  -7.00  & 1.02  \\
17203.750309  &  4.08  & 1.01  \\
17217.748571  &  -3.38  & 1.02
  \enddata
\end{deluxetable}

\begin{figure}
  \includegraphics[width=8.2cm]{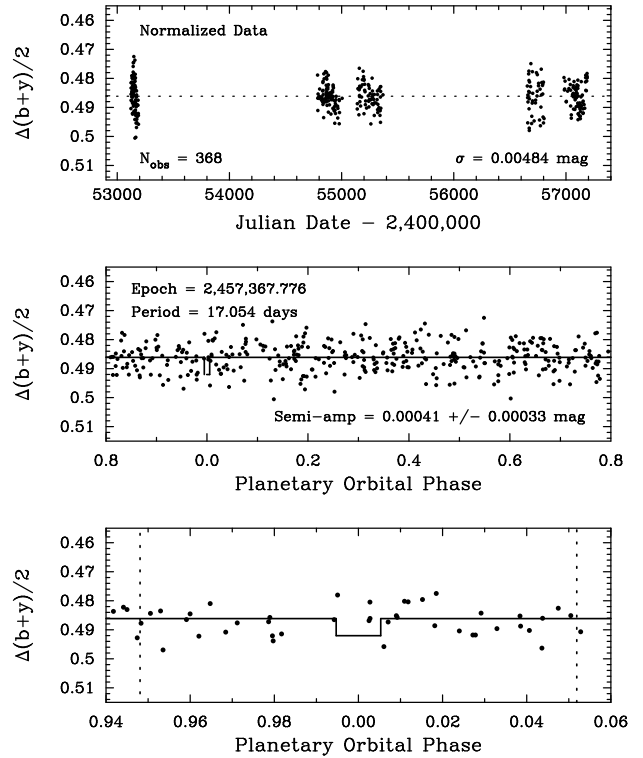}
  \caption{{\it Top}: Nightly photometric observations of HD~99492
    from the 2004, 2009, 2010, 2013, and 2014 observing seasons
    acquired with the T12 0.8~m APT. The final four seasons have been
    normalized so their seasonal means match the 2004 season. {\it
      Middle}: The APT observations phased with the orbital period of
    17.054~days. A sine fit to the phased observations yields a
    semi-amplitude of $0.00041\pm0.00033$~mag. This is consistent with
    the absense of light variability on the radial velocity period and
    also consistent with planetary reflex motion of the star as the
    cause of the RV variations. {\it Bottom}: The APT observations
    within $\pm0.06$ phase units of the predicted transit time. The
    solid curve for the predicted transit shows the predicted
    mid-transit at phase 0.0, the transit depth (0.5\%), and the
    duration ($\pm0.005$ phase units) for a central transit of planet
    b. The vertical dashed lines represent the uncertainty in the time
    of transit. Our photometry shows no evidence for transits but
    cannot rule them out completely.}
  \label{photfig}
\end{figure}

The T12 telescope acquired 368 nightly observations of HD~99492 during
the 2004, 2009, 2010, 2013, and 2014 observing seasons. These data are
plotted against Heliocentirc Julian Date in the top panel of
Figure~\ref{photfig}. The observations are insufficient to detect the
long-term activity cycle described in
Section~\ref{activity}. Therefore, we removed very small
season-to-season variability in HD~99492 and/or its comparison stars
by normalizing the final four observing seasons so their means match
the first season, indicated by the horizonal dotted line in the top
panel. This removal of seasonal variability allows a more sensitive
search for variability that might be due to rotational modulation of
star spots \citep[e.g.,][]{hen13}. \citet{mar05} estimated the
rotation period of HD~99492 to be around 45 days from the Ca II H and
K emission strength.  Our nightly observations scatter about the mean
with a standard deviation 0.00484~mag, somewhat more than the typical
measurement precision. However, Fourier analyses of the complete
normalized data set and the individual observing seasons did not
reveal any significant periodicities between 1 and 100 days that might
correspond to the star's rotation period.

We further examined publicly available photometric data from the {\it
  Hipparcos} satellite to search for evidence of periodicity in the
lightcurve of HD~99492 \citep{per97,van07a}. The data were extracted
from the NASA Exoplanet Archive \citep{ake13}, including 71
measurements spanning a period of 1,062 days and with a standard
deviation of 0.135~mag. Our fourier analysis of the {\it Hipparcos}
did not reveal strong periodicity, with the possible exception of a
minor fourier power at $\sim$15.8~days.

The Keplerian orbital solution in Section~\ref{update} includes an
estimate of $T_c$, the predicted time of mid-transit should the
planetary orbital inclination be suitably close to edge-on. To
determine the remainder of the predicted transit parameters, we
adopted the SME stellar radius from Table~\ref{system} and an
estimated planetary radius of $R_p = 0.52 \ R_J$ using the mass-radius
relationship described by \citet{kan12}. Taking into account the
orbital eccentricity from the Keplerian orbit \citep{kan08}, the
transit probability is 2.8\% and the predicted duration and depth for
a central transit are~0.181 days and 0.54\% respectively.

The APT observations are replotted in the middle panel of Figure
\ref{photfig}. These data are phased with the orbital period and the
predicted transit time shown in Table \ref{system}.  We use
least-squares to fit a sine curve to the data, also phased on the
17.054-day orbital period. This yields a formal semi-amplitude of the
sine curve of just $0.00041\pm0.000033$ mag. The relatively small
amplitude confirms that the observed RV variations are due to the
presence of a planet rather than intrinsic stellar brightness
variations.

The APT observations within $\pm0.06$ phase units of the predicted
transit time are shown in the bottom panel of Figure
\ref{photfig}. The solid curve for the predicted transit signature
includes the predicted mid-transit at phase 0.0, the transit depth
(0.5\%), and transit duration ($\pm0.005$ phase units). The vertical
dashed lines represent the uncertainty in our new time of transit.  We
find no evidence for transits although our data do not rule them out
completely. Monitoring observations were made on the night of 31
January 2016 UT, during a predicted transit, with the T12 APT and with
the 0.6m telescope at Swarthmore's Peter van de Kamp Observatory. The
night was marginally photometric at both sites; again, no evidence for
transits was seen but we are still not able to completely rule them
out.


\section{Conclusions}
\label{conclusions}

The presence of stellar activity presents continuing challenges to
exoplanet detection and characterization. Radial velocity exoplanet
survey targets are usually chosen for their low chromospheric
activity, leading to a bias against activity in the sample of bright
planet-host stars. HD~99491 is an evolved star and has been known to
exhibit chromospheric activity for some time \citep{zar83,wri04}. It
is thus quite interesting to find that the companion star, HD~99492,
exhibits similar behavior over long timescales. It is hoped that
continued photometric monitoring will help to resolve the complete
magentic cycle of the star, such as those found for HD~192263
\citep{dra12}, although the target is very difficult to observe due to
the small angular separation of the binary components.

The update to the parameters for the HD~99492 system presented here
refines the stellar and planetary orbital parameters for the
system. The update shows that the $\sim$5000~day RV signal is due to
stellar activity rather than a planet. However, there are likely other
planets of smaller mass and/or larger separation that lie beneath the
current noise floor. As the exploration of exoplanetary systems forges
onward to ever smaller planets, the careful examination of stellar
activity is becoming more relevant than ever before.


\section*{Acknowledgements}

GWH acknowledges long-term support from Tennessee State University and
the State of Tennessee through its Centers of Excellence program.
This research has made use of the NASA Exoplanet Archive, which is
operated by the California Institute of Technology, under contract
with the National Aeronautics and Space Administration under the
Exoplanet Exploration Program. The results reported herein benefited
from collaborations and/or information exchange within NASA's Nexus
for Exoplanet System Science (NExSS) research coordination network
sponsored by NASA's Science Mission Directorate. The data presented
herein were obtained at the W.M. Keck Observatory, which is operated
as a scientific partnership among the California Institute of
Technology, the University of California and the National Aeronautics
and Space Administration. The Observatory was made possible by the
generous financial support of the W.M. Keck Foundation. The authors
wish to recognize and acknowledge the very significant cultural role
and reverence that the summit of Mauna Kea has always had within the
indigenous Hawaiian community. We are most fortunate to have the
opportunity to conduct observations from this mountain.



\begin{thebibliography}{}

\bibitem[Akeson et al.(2013)]{ake13} Akeson, R.L., Chen, X., Ciardi,
  D., et al. 2013, PASP, 125, 989
\bibitem[Butler et al.(2006)]{but06} Butler, R.P., Wright, J.T.,
  Marcy, G.W., et al. 2006, ApJ, 646, 505
\bibitem[Carolo et al.(2014)]{car14} Carolo, E., Desidera, S.,
  Gratton, R., et al. 2014, A\&A, 567, A48
\bibitem[Demarque et al.(2004)]{dem04} Demarque, P., Woo, J.-H., Kim,
  Y.-C., Yi, S.K. 2004, ApJS, 155, 667
\bibitem[Deming et al.(1987)]{dem87} Deming, D., Espenak, F.,
  Jennings, D.E., Brault, J.W., Wagner, J. 1987, ApJ, 316, 771
\bibitem[Desort et al.(2007)]{des07} Desort, M., Lagrange, A.-M.,
  Galland, F., Udry, S., Mayor, M. 2007, A\&A, 473, 983
\bibitem[Dragomir et al.(2012)]{dra12} Dragomir, D., Kane, S.R.,
  Henry, G.W., et al. 2012, ApJ, 754, 37
\bibitem[Dravins(1985)]{dra85} Dravins, D. 1985, in Stellar Radial
  Velocities, Proceedings of IAU Colloquium No. 88, ed. A.G.D. Philip
  \& D.W. Latham, 311-320
\bibitem[Hatzes(2013)]{hat13} Hatzes, A.P. 2013, ApJ, 770, 133
\bibitem[Hatzes et al.(2015)]{hat15} Hatzes, A.P., Cochran, W.D.,
  Endl, M., et al. 2015, A\&A, 580, 31
\bibitem[H\'ebrard et al.(2014)]{heb14} H\'ebrard, \'E.M., Donati,
  J.-F., Delfosse, X., Morin, J., Boisse, I., Moutou, C., H\'ebrard,
  G. 2014, MNRAS, 443, 2599
\bibitem[Henry(1999)]{hen99} Henry, G.W. 1999, PASP, 111, 845
\bibitem[Henry et al.(2013)]{hen13} Henry, G.W., Kane, S.R., Wang,
  S.X., et al. 2013, ApJ, 768, 155
\bibitem[Hinkel et al.(2014)]{hin14} Hinkel, N.R., Timmes, F.X.,
  Young, P.A., Pagano, M.D., Turnbull, M.C. 2014, AJ, 148, 54
\bibitem[Howard et al.(2009)]{how09} Howard, A. W., Johnson, J.A.,
  Marcy, G.W., et al. 2009, ApJ, 696, 75
\bibitem[Isaacson \& Fischer(2010)]{isa10} Isaacson, H., Fischer,
  D. 2010, ApJ, 725, 875
\bibitem[Kane \& von Braun(2008)]{kan08} Kane, S.R., von Braun,
  K. 2008, ApJ, 689, 492
\bibitem[Kane et al.(2009)]{kan09} Kane, S.R., Mahadevan, S., von
  Braun, K., Laughlin, G., Ciardi, D.R.  2009, PASP, 121, 1386
\bibitem[Kane \& Gelino(2012)]{kan12} Kane, S.R., Gelino, D.M. 2012,
  PASP, 124, 323
\bibitem[Lodders et al.(2009)]{lod09} Lodders, K., Plame, H., Gail,
  H.-P. 2009, Landolt-B{\"o}rnstein - Group VI Astronomy and
  Astrophysics Numerical Data and Functional Relationships in Science
  and Technology Volume 4B: Solar System. Edited by J.E. Tr{\"u}mper,
  4B, 44
\bibitem[Marcy \& Butler(1992)]{mar92} Marcy, G.W., Butler, R.P. 1992,
  PASP, 104, 270
\bibitem[Marcy et al.(2005)]{mar05} Marcy, G.W., Butler, R.P., Vogt,
  S.S., et al. 2005, ApJ, 619, 570
\bibitem[Meschiari et al.(2011)]{mes11} Meschiari, S., Laughlin, G.,
  Vogt, S.S., et al. 2011, ApJ, 727, 117
\bibitem[Moutou et al.(2011)]{mou11} Moutou, C., Mayor, M., Lo Curto,
  G., et al. 2011, A\&A, 527, A63
\bibitem[Noyes et al.(1984)]{noy84} Noyes, R.W., Hartmann, L.W.,
  Baliunas, S.L., Duncan, D.K., Vaughan, A.H. 1984, ApJ, 279, 763
\bibitem[Perryman et al.(1997)]{per97} Perryman, M.A.C., Lindegren,
  L., Kovalevsky, J., et al. 1997, A\&A, 323, L49
\bibitem[Petigura \& Marcy(2011)]{pet11} Petigura, E.A., Marcy,
  G.W. 2011, ApJ, 735, 41
\bibitem[Queloz et al.(2001)]{que01} Queloz, D., Henry, G.W., Sivan,
  J.P., et al. 2001, A\&A, 379, 279
\bibitem[Robertson et al.(2013)]{rob13} Robertson, P., Endl, M.,
  Cochran, W.D., MacQueen, P.J., Boss, A.P. 2013, ApJ, 774, 147
\bibitem[Robertson et al.(2014)]{rob14} Robertson, P., Mahadevan, S.,
  Endl, M., Roy, A. 2014, Science, 345, 440
\bibitem[Robertson et al.(2015)]{rob15} Robertson, P., Roy, A.,
  Mahadevan, S. 2015, ApJ, 805, L22
\bibitem[Saar \& Donahue(1997)]{saa97} Saar, S.H., Donahue, R.A. 1997,
  ApJ, 485, 319
\bibitem[Santos et al.(2010)]{san10} Santos, N.C., Gomes da Silva, J.,
  Lovis, C., Melo, C. 2010, A\&A, 511, A54
\bibitem[Takeda et al.(2007)]{tak07} Takeda, G., Ford, E.B., Sills,
  A., et al. 2007, ApJS, 168, 297
\bibitem[Valenti et al.(1995)]{val95} Valenti, J.A., Butler, R.P.,
  Marcy, G.W. 1995, PASP, 107, 966
\bibitem[Valenti \& Piskunov(1996)]{val96} Valenti, J.A., Piskunov,
  N. 1996, A\&AS, 118, 595
\bibitem[Valenti \& Fischer(2005)]{val05} Valenti, J.A., Fischer,
  D.A. 2005, ApJS, 159, 141
\bibitem[Valenti et al.(2009)]{val09} Valenti, J.A., Fischer, D.,
  Marcy, G.W., et al. 2009, ApJ, 702, 989
\bibitem[van Leeuwen(2007a)]{van07a} van Leeuwen, F. 2007a, Hipparcos,
  the New Reduction of the Raw Data, Astrophys. Space Sci. Lib., 350
\bibitem[van Leeuwen(2007b)]{van07b} van Leeuwen, F. 2007b, A\&A, 474,
  653
\bibitem[Vogt et al.(1994)]{vog94} Vogt, S.S., Allen, S.L., Bigelow,
  B.C., et al. 1994, Proc. SPIE, 2198, 362
\bibitem[Wang et al.(2012)]{wan12} Wang, S.X., Wright, J.T., Cochran,
  W., et al. 2012, ApJ, 761, 46
\bibitem[Wilson(1968)]{wil68} Wilson, O.C. 1968, ApJ, 153, 221
\bibitem[Wright et al.(2004)]{wri04} Wright, J.T., Marcy, G.W.,
  Butler, R.P., Vogt, S.S. 2004, ApJS, 152, 261
\bibitem[Wright(2005)]{wri05} Wright, J.T. 2005, PASP, 117, 657
\bibitem[Wright et al.(2008)]{wri08} Wright, J.T., Marcy, G.W.,
  Butler, R.P., et al. 2008, ApJ, 683, L63
\bibitem[Wright \& Howard(2009)]{wri09} Wright, J.T., Howard,
  A.W. 2009, ApJS, 182, 205
\bibitem[Zarro(1983)]{zar83} Zarro, D.M. 1983, ApJ, 267, L61

\end{thebibliography}
\end{document}